# DETECTION OF SEVERAL DAEMON POPULATIONS IN EARTH-CROSSING ORBITS

E.M.Drobyshevski, M.V.Beloborodyy, R.O.Kurakin, V.G.Latypov, and K.A.Pelepelin

*A.F.Ioffe Physical-Technical Institute, Russian Academy of Sciences,
194021 St.Petersburg, Russia
(E-mail: emdrob@pop.ioffe.rssi.ru)*

## Abstract

Experiments on scintillator-based detection of negative daemons, DArk Electric Matter Objects, representing Planckian supermassive ($\sim 2 \cdot 10^{-5}$ g) particles, whose population has been detected in March 2000 to populate near-Earth, almost circular, heliocentric orbits (NEACHOs), are being continued. The NEACHO objects hit the Earth with a velocity ~10-15 km/s. The results of these and new experiments (April through June 2001) are now processed taking into account the difference in scintillation signal shape depending on the magnitude and sign of the velocity of the daemons crossing our detector, which was purposefully made asymmetric with respect to the up/down direction of flight. The data accumulated during the time of the experiment and processed in this way reveal also the presence of (1) a high-velocity (~35-50 km/s) daemon population whose objects can be related to a population in the Galactic disk and/or that in strongly elongated, Earth-crossing heliocentric orbits (SEECHOs), as well as (2) a low-velocity (~3-10 km/s) population in geocentric Earth-surface-crossing orbits (GESCOs), whose objects traverse repeatedly the Earth to suffer a decrease in velocity by ~30-40% in a month in the process.

An evolutionary relation between all these three (four?) populations is discussed. Assumptions concerning their manifestations in further observations are put forward.

An analysis of possible interaction processes of daemons, which may have different velocities and directions of motion, with the detector components [ZnS(Ag) scintillator layers, 0.3-mm thick tinned-iron sheets etc.] on the atomic (emission of Auger electrons) and nuclear (nucleon evaporation from a nucleus excited in the capture and, subsequently, the decay of its protons) levels has permitted estimation of some characteristic times. In particular, the decay time of a daemon-containing proton was found to be ~1 μs.

## 1. Introduction. Description of the Detector and the Results of the First Experiments

We believe a substantial part of the Galactic disk DM to be made up of DArk Electric Matter Objects, daemons (Drobyshevski 1996). They are relic elementary black holes moving with astronomical velocities and having mass of a Planckian scale $M \approx 2 \cdot 10^{-5}$ g. It may be conjectured that the evolution of their ensemble with initial spectrum close to δ-function had favored at the time an appearance of the barion asymmetric visible matter (Barrow et al. 1992) and further the formation of the large-scale structure of the Universe as we know it, e.g. quasars, (super)clusters of galaxies with their (active) nuclei and spiral arms, etc. The Planckian mass allows these black holes to have an electric charge up to $Z \approx 10$ (Markov 1965). Negative daemons are nuclear-active particles. In building up inside the Sun, they catalyze proton fusion, which may account for the observed solar energetics and the deficiency of the emitted neutrinos (Drobyshevski 1996).

As the Sun captures daemons by slowing down their motion, some of them, acted upon by the gravitational perturbations by the Earth, should populate also strongly elongated Earth-crossing heliocentric orbits (SEECHOs) (Drobyshevski 1997a, 2000b). A certain part of them transfers in due time to near-Earth, almost circular heliocentric orbits (NEACHOs), from which they can now be transferred by the Earth into geocentric orbits crossing its surface (Drobyshevski 2000c, 2001).

Estimates show (Drobyshevski 1997a) that the volume concentration of daemons in SEECHOs exceeds that in the Galactic disk by a factor $\sim 10^3$-$10^5$. Therefore, in contrast to the standard practice adopted when searching for the Galactic halo DM objects moving with velocities of 200-300 km/s (see, e.g. Bauer 1998, Spooner 1998), our experiments were aimed from the very beginning at looking for a relatively low-velocity (35-50 km/s), but high-density population in SEECHOs. We planned to make use of the active interaction of negative daemons with atomic nuclei, including the decay of a daemon-containing proton in a time $\Delta t_{ex} \sim 10^{-7}$-$10^{-6}$ s (Drobyshevski 2000a,b).

We developed a simple four-module detector (Drobyshevski 2000c, 2001). Each module contains two transparent polystyrene plates arranged horizontally one above the other at a distance of 7 cm (Fig. 1). They are 4 mm thick and measure $50 \times 50$ cm². The plates were coated on the downside with a ~3.5 mg/cm² layer of ZnS(Ag) powder with an average grain size ≈12 μm. Each plate was viewed by its PM tube. The plates were mounted at the center of cubic box, 51 cm on a side, made of 0.3-mm thick iron sheet faced on both sides with 2 μm-thick tin layer.

The upper cover of the box was of black paper. As will be seen later, all the components of the detector served to various degrees as its active elements.

The signals from each pair of the PM tubes were fed into a dual-trace digital storage oscillograph, which recorded them with a lead/delay of ±100 μs when triggered by a signal from the upper PM tube. If signals *shifted* relative to one another appeared on both traces, they were fed into a computer for subsequent processing.

The experiments were originally targeted to detect the SEECHO population. We expected to find signals shifted by $\Delta t = 7$ cm/(35-50 km/s) ≈ 1.5-2 μs. However the 700-h exposure made in March 2000 revealed an excess of signals at a level of 3.8σ within the interval $+20 < \Delta t < +40$ μs, which corresponds to a downward flux of some objects (Fig. 2c); the most significant events were those having on the first trace fairly long signals characteristic of scintillations caused by heavy nonrelativistic particles of the type of α-particles (Heavy Particle Scintillations - HPS) (Drobyshevski 2000d). Retrospectively, we think we were very lucky to stumble, during the very first month of observations, across this significant feature. If we had not recognized it as such, we would hardly have had stimulus and persistence enough to continue our observations.

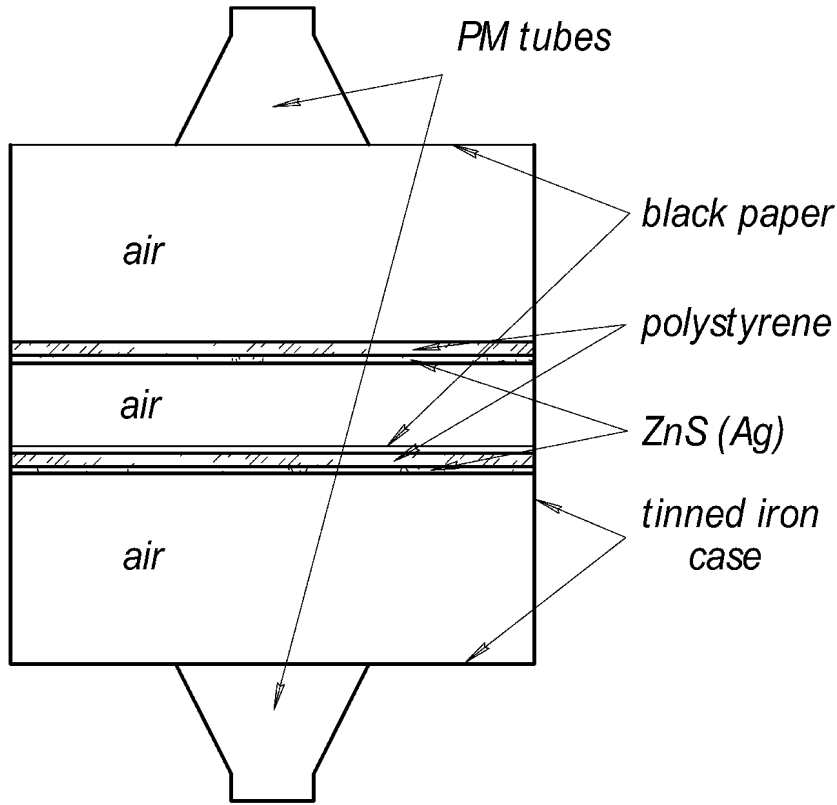

Fig.1. Schematic of the scintillator module. The tinned iron cubic box is 51 cm on side, spacing between ZnS(Ag) coatings is 7 cm.

The position of the peak at $\Delta t \approx +30$ μs permitted us to immediately estimate the flux velocity as $V \approx 7$ cm/30 μs $\approx 2.5$ km/s, which, taking into account the possible inclination of the trajectories, could be raised to ~5 km/s. There were grounds to conclude that we have detected particles trapped by the Earth from the SEECHO population into geocentric orbits crossing the Earth's surface (Drobyshevski 2000c, 2001).

A more comprehensive analysis of possible implications of daemon interaction with the box wall material, combined with the narrowness of the peak observed at $\Delta t = +30\pm10$ μs, suggested, however, that the weaker and shorter lower scintillations, primarily of the NLS type (Noise-Like Scintillations), are due to numerous Auger electrons with energies of up to 0.6-1 MeV emitted in the capture of Sn and Fe atoms in the lower cover of the box, which corresponds to $V = (22$ cm $+ 7$ cm$)/30$ μs $= 10$ km/s. Taking into account the trajectory inclinations within a solid angle $\approx 4\pi/6 = 2$ ster cut out by the side walls of the upper half of the box yields $V = 10\text{-}15$ km/s. Such a velocity is characteristic of objects falling on the Earth from NEACHOs (Drobyshevski 2000d).

Sec. 2 will consider in more detail the processes accompanying the passage of a daemon through the components of our system.

Another problem which faced us was the strong seasonal variations in the daemon flux.

We have to confess that initially we had believed the weak (~7%) seasonal variations in the WIMP flux to be peculiar only to the Galactic halo population with $V \approx 200\text{-}300$ km/s.

We knew certainly that the Sun's velocity relative to the nearest stars is ~20 km/s, but assumed as a working hypothesis that the velocity relative to the nearest daemon background in the Galactic disk is, on the average, negligible, and small compared to the random velocity of the objects in this background. We did not intend to reach an accuracy in our measurements better than 10%. As a result, on recording the above-mentioned maximum and carrying out a series of tests of the reliability of our measuring equipment we were surprised to find the results obtained a couple of months later (a typical series of observations lasts about a month) to very poorly reproduce the March 2000 data. This initiated a large number of improvements of the equipment and of its repeated tests, additional nonsystematic experiments, attempts at devising another interpretation of their results (e.g. Drobyshevski 2000c,d,e; 2001), etc. At the same time we made gradual progress in understanding the character of the interaction of daemons with matter and their celestial mechanics, and developed new promising techniques of data treatment (see Sec. 2). Therefore, when, at the very end of March, 2001, we again started systematic observations and did not reveal a maximum as significant as the one found one year and a month before, we were not very much discouraged. A more sophisticated treatment of both old and new series of data accumulated during the period from April through June 2001 revealed new types of daemon populations in the Solar system (Sec. 3) and their evolution with time (Sec. 4). This provided supportive evidence for the correctness of the strategy chosen by us, as well as the validity of the main ideas concerning the generic relation between the various daemon populations and their interaction with matter. As a result, we have been able to estimate the characteristic times of some processes occurring on the atomic and subnuclear levels, and, in particular, have obtained a more precise estimate of the decay time of the daemon-containing proton (Sec. 3). We certainly are aware of the fact that many important details of these processes still remain unclear and need a more comprehensive investigation (Sec. 3).

## 2. Detector Asymmetry: Its Reasons and Merits

As already mentioned, our detecting system was purposefully made asymmetric with respect to the up/down direction in order not only to reveal the very existence of daemons but to broaden the scope of information concerning their properties and interaction with matter. The asymmetry was provided by depositing the ZnS(Ag) layer only on the downside of the polystyrene plates and by the lack of the top iron cover on the box. It was assumed that the upward-moving daemons would certainly be poisoned by heavy Sn or Fe nuclei, thus changing dramatically their properties, which would not happen with a downward flux, etc.

Table 1 gives some idea of the details of such processes. It presents path lengths $l$ of a super-massive particle with a negative charge $Z_{eff}e = 1e$ before it has recaptured an atomic nucleus, calculated for different substances [polystyrene, air, ZnS(Ag), iron, tin] at three characteristic velocities $V = 5$, 15, and 45 km/s. The choice of such a low value of $Z_{eff}$ for the daemon is suggested by the totality of our experimental results, if for no other reason than that the characteristic times of flight in our system are, on the whole, fairly short (or, at most, of the same order of magnitude) compared to the decay times of daemon-containing nuclei (and of the protons they are made of). It is not inconceivable that sometimes one should adopt $Z_{eff} = 2$ or even $Z_{eff} = Z \approx 10$. It appears that large values (up to $Z_{eff} \approx 10$) may possibly be realized for a long enough time only in large evacuated volumes. The cross section for a (unexcited) nucleus with charge $Z_n$ and mass $Am_p$, taking into account its displacement by a negative projectile, was determined from the expression (Drobyshevski 2000b)

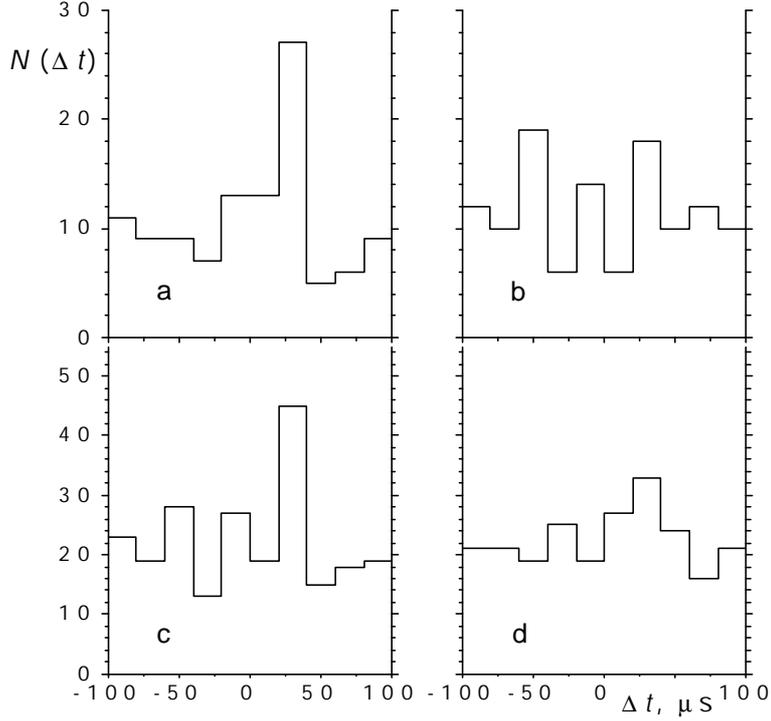

Fig.2. 24 February – 27 March, 2000, statistics for shifted scintillations of upper and lower scintillators. Only Heavy-Particle Scintillation (HPS) events at the upper scintillator, which activate the oscilloscopic sweep, are taken into account.

a – the "wide" HPS distribution $N_w(\Delta t)$;

b – the "narrow" HPS distribution $N_n(\Delta t)$ (for definition of $N_n(\Delta t)$ and $N_w(\Delta t)$ see Fig. 3);

c – the combined $N(\Delta t) = N_w+N_n$ demonstrating a sharp peak in $+20<\Delta t<40$ μs bin;

d – the "inverted" combined distribution $N_w(\Delta t)+N_n(-\Delta t)$ which sets off dependence of the HPS width on the up/down direction of the daemon motion; this dependence is most prominent for $\pm 20$ μs (see Figs. 4d also).

$$\boldsymbol{s} = \boldsymbol{s}_0 \frac{2Z_n e^2}{m_p R_0 A^{4/3}} \cdot \frac{Z_{eff}}{V^2}, \qquad (1)$$

where $\boldsymbol{s}_0 = \pi R_0^2 A^{2/3}$ and $R_0 = 1.2 \cdot 10^{-13}$ cm.

To assist in orienting oneself in the characteristic values of the quantities involved and to make possible their comparison, Table 1 presents also the dimensions $l$ and the corresponding times of flight $t_l$ for our system, as well as the Knudsen numbers $K = \boldsymbol{l}/l$.

It would seem that in crossing the tinned iron sheet of the box, a daemon should inevitably capture an iron nucleus, but slow daemons with $V \leq 10\text{-}15$ km/s will be poisoned by Sn nuclei before that. By contrast, when daemons cross the ZnS(Ag) layer, a noticeable fraction of them, having velocities $V \geq 35$ km/s, may pass through without having interacted with a Zn or S nucleus at all.

The asymmetry of the system should affect also the shape of the scintillations. It appears nearly obvious, that a nucleus excited when captured in the ZnS(Ag) layer will not be able to evaporate all of its nucleons in the de-excitation during the time it takes to cross the thin luminophor layer. Therefore the HPS produced by the nucleons evaporating in the polystyrene as the daemon moves upward should be, on the average, shorter and weaker than that generated in its downward flight.

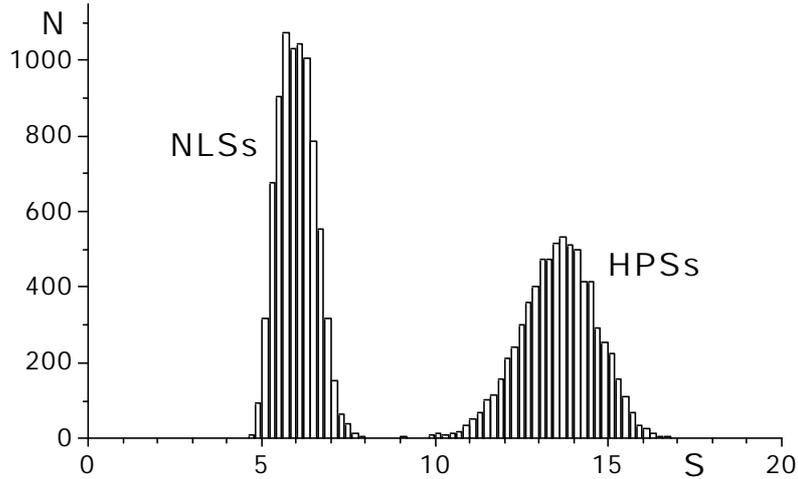

Fig.3. Typical distribution of the noise-like (light-particle) scintillations (NLSs) and HPSs vs. S – the signal-amplitude normalized area swept by the oscilloscopic trace of the scintillation. These distributions can be approximated by Gaussians. The LPS and HPS distributions are well distinctive by their S's. The "narrow" HPSs and "wide" HPSs lie at left and right parts of the HPS Gaussian, correspondingly.

Fig. 3 displays our data on the typical distributions of NLSs and HPSs plotted as a function of the area bounded by their oscilloscopic traces and normalized to the amplitude of the signal. (In experiments of 2001, we no longer insert an inductance into the PM tube load circuit as it was done earlier (Drobyshevski 2000c,d,e); now the signal is picked off from a resistance $R = 30$ k$\Omega$ connected with the PM anode through $R = 1.5$ k$\Omega$; the capacity of the cable feeding the PM tube output signal to the oscillograph is 110 pF.) The NLS and PHS distributions are seen to be clearly separated. They can be well fitted by Gaussians. It is also evident that both distributions are produced primarily by background signals. Nevertheless, in view of the above remark on a possible effect of the direction in which a daemon crosses the ZnS(Ag) layer on the scintillation shape it appeared only natural to compare the parts of the distribution $N(\Delta t) = N_n(\Delta t) + N_w(\Delta t)$ constructed for the events contained in the left-hand ($N_n$, "narrow" HPSs) and the right-hand ($N_w$, "wide" HPSs) wings of the HPS Gaussian. This comparison (see Figs. 2a,b and 4a,b) yielded very important information both on the processes accompanying the nucleus capture by a daemon and on the existence of fairly distinct daemon populations and their possible generic relation.

Table 1. Parameters of the detector components and quantities characterizing their interactions with daemons at three typical velocities ($V$ = 5, 15, and 45 km/s). $l$ is the component size and $t_l$ is time of its traversing by daemon; $\mathbf{\mathit{l}}$ is the mean free path needed for a daemon with $Z_{\text{eff}} = 1$ (the most probable case) to capture a nucleus, $\mathbf{\mathit{t_l}}$ is the corresponding time (to calculate $\mathbf{\mathit{l}}$ and $\mathbf{\mathit{t_l}}$, the cross-section defined by Eq.1 is used). $K = \mathbf{\mathit{l}}\, Z_{\text{eff}}/l$ is the Knudsen number.

| | $Z_n$ | A | $r$ [g/cm$^3$] | $l$ [cm] | $t_l$ [μs] | | | $\mathbf{\mathit{l}} \cdot Z_{\text{eff}}$ [cm] | | | $\mathbf{\mathit{t_l}} \cdot Z_{\text{eff}}$ [ns] | | | $K = \mathbf{\mathit{l}}\, Z_{\text{eff}}/l$ | | |
|---|---|---|---|---|---|---|---|---|---|---|---|---|---|---|---|---|
| V [km/s] | | | | | 5 | 15 | 45 | 5 | 15 | 45 | 5 | 15 | 45 | 5 | 15 | 45 |
| H | 1 | 1 | (0.082) | | | | | 4.9·10⁻⁵ | 4.4·10⁻⁴ | 4.0·10⁻³ | 0.1 | 0.29 | 0.89 | | | |
| C | 6 | 12 | (0.978) | | | | | 5.2·10⁻⁵ | 4.7·10⁻⁴ | 4.2·10⁻³ | 0.1 | 0.31 | 0.93 | | | |
| poly-styrene | CH | | 1.06 | 0.4 | 0.8 | 0.27 | 0.09 | 2.5·10⁻⁵ | 2.3·10⁻⁴ | 2.0·10⁻³ | 0.05 | 0.15 | 0.44 | 6.3·10⁻⁵ | 5.6·10⁻⁴ | 5·10⁻³ |
| N | 7 | 14 | (0.00091) | | | | | 5.1·10⁻² | 4.6·10⁻¹ | 4.1 | 102 | 307 | 910 | | | |
| O | 8 | 16 | (0.00028) | | | | | 0.18 | 1.6 | 14.5 | 360 | 1070 | 3200 | | | |
| air | | | 0.0012 | 7 | 14 | 4.7 | 1.6 | 0.04 | 0.36 | 3.2 | 80 | 240 | 710 | 5.7·10⁻³ | 5.3·10⁻² | 0.45 |
| | | | | 22 | 44 | 14.7 | 4.9 | | | | | | | 1.8·10⁻³ | 1.6·10⁻² | 0.15 |
| S | 16 | 32 | (1.35) | | | | | 5.8·10⁻⁵ | 5.2·10⁻⁴ | 4.7·10⁻³ | 0.12 | 0.35 | 1.04 | | | |
| Zn | 30 | 65 | (2.74) | | | | | 5.1·10⁻⁵ | 4.5·10⁻⁴ | 4.1·10⁻³ | 0.10 | 0.30 | 0.91 | | | |
| ZnS(Ag) | | | 4.09 | 10⁻³ | 2·10⁻³ | 0.7·10⁻³ | 2.2·10⁻⁴ | 2.7·10⁻⁵ | 2.4·10⁻⁴ | 2.2·10⁻³ | 0.054 | 0.16 | 0.49 | 2.7·10⁻² | 0.24 | 2.17 |
| Fe | 26 | 56 | 7.9 | 0.03 | 0.6 | 0.02 | 0.007 | 1.6·10⁻⁵ | 1.4·10⁻⁴ | 1.3·10⁻³ | 0.032 | 0.093 | 0.29 | 5.3·10⁻⁴ | 4.7·10⁻³ | 4.3·10⁻² |
| Sn | 50 | 119 | 7.3 | 2·10⁻⁴ | 4·10⁻⁴ | 1.3·10⁻⁴ | 4.4·10⁻⁵ | 3.1·10⁻⁵ | 2.8·10⁻⁴ | 2.5·10⁻³ | 0.062 | 0.19 | 0.56 | 0.16 | 1.4 | 12.5 |

### 3. Daemon Populations (Families)

*3.1. The fast component of the daemon flux and estimation of $\Delta t_{ev}$, the decay time of a daemon-containing proton*

The first thing that strikes the eye in the *a* and *b* histograms of Fig. 2 and Fig. 4 is the inverse ratios of the numbers of events contained in the two central bins (±20 μs) for the "narrow", $N_n(\Delta t)$, and "wide", $N_w(\Delta t)$, parts of the HPS $N(\Delta t)$ distribution. The upward flux ($-20 \leq \Delta t \leq 0$ μs) generates a larger number of "narrow" than "wide" events compared to the downward-moving flux ($0 < \Delta t < +20$ μs). It is just what we expected to find (Drobyshevski 2000c,d; 2001) (see also Sec. 2) by assuming that the evaporative de-excitation time of a captured nucleus, $t_{ev}$, exceeds noticeably the daemon flight time $t_l$ through the ~10-μm thick ZnS(Ag) layer, and that particle evaporation from an upward moving nucleus ends in the polystyrene (a useful note: the mean free path of a 1MeV α-particle in polystyrene ~20 μm). Whence it follows (see Table 1) that $t_{ev} > \sim 2 \cdot 10^{-10}$ s (if a scintillator nucleus is captured at this velocity at all, see below). A downward-moving daemon-containing nucleus exiting ZnS(Ag) into air causes a fairly wide, "normal" HPS.

The events falling in the $-20 < \Delta t < +20$ μs bins are produced by daemons moving with velocities $\sim 15 \leq V \leq 50$ km/s, if they are calculated based on the distance from the lower box cover to the upper ZnS(Ag) layer. Taking into account the trajectory inclinations may increase *V* by a factor 1.5-2, raising the lower limit of *V*, accordingly, at least up to 20-30 km/s. As for the upper bound on the velocity, a trial exclusion of the few events corresponding to $-2 \leq \Delta t \leq +2$ μs from the central bins did not result in any significant change.

The data available to date permit a conclusion that we are observing objects populating SEECHOs ($V \approx 35$-$50$ km/s) and/or those of the interstellar background, i.e., of the Galactic disk population, if the Sun moves relative to it with $V \approx 10$-$20$ km/s (then, taking into account the Earth's orbital motion, we come again to $V_{max} \approx (10\text{-}20) + V_{\oplus orb} \approx 35\text{-}50$ km/s). Measurements spanning even only one year would apparently be enough to discriminate between these two populations with a good confidence. Thus, we may expect that in about half a year after the observations reported here, when $V_{\oplus orb}$ changes direction relative to the incoming flux from the Galactic disk, there will be a certain deficiency of events in the $-20 < \Delta t < +20$ μs interval.

As seen from Table 1, for $V \approx 35$-$50$ km/s a daemon with $Z_{eff} = 1$ crosses the 2 μm tin layer practically without Sn nucleus capture, to capture with a ~100% probability a Fe nucleus in the 0.3-mm thick iron sheet. The probability to capture a S or Zn nucleus in the ZnS(Ag) layer is about 50%.

The fact that the upward flux produces in the $-20 < \Delta t < +20$ μs interval always more "narrow" HPSs, and the downward flux, more "wide" HPSs, permits a conclusion that in the first case in the time $\Delta t \leq 20$ μs elapsed between the capture of a Fe nucleus in the lower box cover and its entering the upper scintillator layer, some of about $\sim Z_n - 9 = 17$ protons are evaporated from the excited Fe nucleus, while the remainder of these 17 protons become subsequently "digested" by the daemon it contains.

This estimate is obtained also for the downward flux of daemons, which capture S or Zn in the upper scintillator layer and acquire, before reaching the lower box cover in $\Delta t \leq 20$ μs, an effective charge $Z_{eff} \geq 1$, i.e., an ability to capture in it a Fe nucleus, with subsequent emission of energetic (up to ~0.6-1 MeV) Auger electrons, which penetrate through the ~0.3 mm iron sheet and initiate (together with the δ-electrons) a scintillation in the lower ZnS(Ag) layer.

Bearing in mind that the binding energy of a "pure" daemon ($Z_{eff} = 10$) with a nucleus $Z_n \approx A/2 = Z_{eff} = 9$ is as high as $W \approx 1.5 \cdot Z_{eff} A^{2/3} \sim 103$ MeV, and that with an iron nucleus $Z_n = 26$ this is $W \sim 210$ MeV, we obtain ~105 MeV for the energy expended immediately to heat the iron nucleus newly captured in place of the nucleus with $Z_n = 9$. If the average nucleon binding energy in a nucleus ≈7.5 MeV, this energy is high enough to evaporate, in $t_{ev} \ll 20$ μs, ~ 14 nucleons (or 3-4 α-particles). As a consequence, the de-excited Fe nucleus transforms to a nucleus with $Z_n \leq 19$-20 (approximately the same figures apply to Zn; the nucleus left over after the evaporation of nucleons in S in $t_{ev}$ will have $Z_n \approx 14$). After this, consecutive daemon decay of the "excess" $Z_n - Z_{eff} = 20 - 9 = 11$ protons takes place in $\Delta t \leq 20$ μs. This yields $\Delta t_{ex} \leq 20/11 \approx 1.8$ μs for the decay of a daemon-containing proton.

*3.2. Daemons in near-Earth almost-circular heliocentric orbits (NEACHOs)*

The $N(\Delta t)$ distribution accumulated in March 2000 exhibits a fairly narrow maximum at $\Delta t \approx +30\pm10$ μs at a confidence level of ~99.9% (Fig. 2c). There is no symmetric counterpart to it for $\Delta t < 0$. Taking as a reference the distance $l = 29$ cm between the upper scintillator and the lower box cover, we obtain (with due account of trajectory inclination within ~2 ster) a velocity $V \approx 10$-15 km/s. This figure is in a good agreement with the value expected for NEACHO objects falling on the Earth (Drobyshevski 2000d). The evolution of the heliocentric and other daemon populations is covered in more detail in Sec. 4.

At such a velocity, the probability for a daemon to capture a Sn nucleus in the box walls is close to unity (see Table 1), which accounts for the absence of signals due to the corresponding upward flux, as well as for the narrowness itself of the distribution ($\Delta t \approx +30\pm10$ μs), because the side walls of the upper half of the box cut out from above a solid angle ≈ 2 ster.

For $Z = 10$, the energy released in the capture of a Sn nucleus $W \sim 320$ MeV, which leaves ~215 MeV for the excitation of the Sn nucleus and evaporation of its nucleons (for the initial $Z_{eff} = 9$). This energy is large enough to evaporate ~28 nucleons, with about 40 protons left in the remaining nucleus. Whence it follows that $\Delta t_{ex} \geq (30$ μs$)/(40 - 9) \approx 1$ μs. Taking into account the data presented in Sec. 3.1, we come to $1.8 \geq \Delta t_{ex} \geq 1$ μs.

In these estimates we disregard, on the one hand, the possibility of excitation energy being removed by γ-radiation and, on the other, of the nucleus ejecting nucleon clusters.

Note also that the $+20 < \Delta t < 40$ μs bin contains, to within a statistical error, approximately equal numbers of the wide and narrow HPSs.

*3.3. Low-Velocity Orbit Population*

*3.3.1. Evidence for the existence of a geocentric population*

It could be assumed that all daemons belong to two (possibly, three) above-mentioned populations, and that beyond the $\Delta t = \pm 40$ μs interval the $N(\Delta t)$ population contains background signals only. A comparison of the $N_w(\Delta t)$ and $N_n(\Delta t)$ distributions in Figs. 4.1a,b shows, however, that the wide HPSs lie, as a rule, within $-80 \leq \Delta t \leq -40$ μs, and the narrow ones, within $+80 \leq \Delta t \leq +40$ μs. If these were purely background signals, there would be no such correlation with $\Delta t$. Thus, we have to admit that there is one more daemon population, whose upward flux initiates the wide HPSs, and the downward flux, the narrow ones. This situation is the inverse to that with the SEECHO objects.

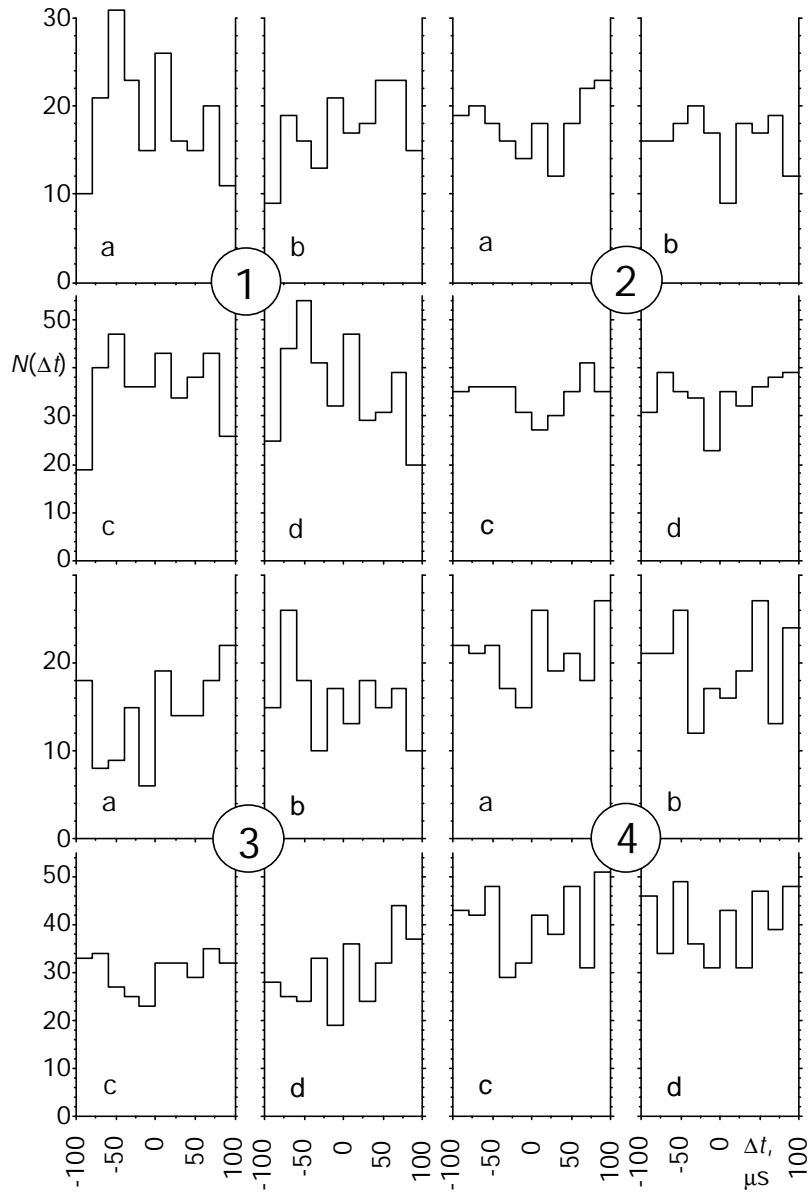

Fig.4. The same as Fig.2 for different time intervals in 2001:
(1) 30 March - 25 April (590 h·m$^2$);
(2) 25 April - 24 May (660 h·m$^2$);
(3) 24 May - 15 June (700 h·m$^2$);
(4) 15 June - 30 June (760 h·m$^2$).
Note that from 6 June and on the second 4-module detector was put into operation, in parallel with the first one (the total detector area become 2 m$^2$).

The April 2000 $N(\Delta t)$ distribution (Fig. 4.1c) exhibits two, nearly symmetric maxima at $\Delta t \approx -(40\text{-}80)$ μs and $\Delta t \approx +(40\text{-}80)$ μs, which corresponds to a velocity of 5-10 km/s. It appears only natural to assume that we observe here indeed a population in geocentric Earth-surface-crossing orbits (GESCOs). We believed earlier that it is the objects with $+20 < \Delta t < +40$ μs that are due to this population (Drobyshevski 2000c, 2001).

For velocities ≤10 km/s, the 2-μm tin layer becomes, as we have seen, opaque to the capture of a Sn nucleus. The maximum at $\Delta t < -40$ μs suggests that by this time the upward moving daemons disintegrate a Sn nucleus to the extent where their charge becomes $Z_{eff} = 1$. Whence it follows, in accordance with the calculations made at the end of Sec. 3.2, that $\Delta \boldsymbol{t}_{ex} \leq (40\ \mu s)/(40 - 9) \approx 1.3$ μs. Recalling the estimates made in Secs. 3.2 and 3.1, we come to $\Delta \boldsymbol{t}_{ex} \approx 1\text{-}1.3$ μs.

This result follows directly from the difference in the behavior of $N(\Delta t)$ observed to occur as $\Delta t$ crosses the value –40 μs. It certainly needs refining by taking into account the possible emission of γ-quanta and nucleon clusters, more accurate calculation of the binding energies etc. One may also conclude, in particular, that the 7-cm gap between the scintillators can be adopted as a reference only for the low-velocity daemons (≤5 km/s), i.e., when their time of flight through this gap is longer than the time the daemon needs to shed off (at least partially, to $Z_n < Z$) the nucleus it had captured in the scintillator (see the end of Sec. 3.1 and Sec. 4).

*3.3.2. On the dependence of the visible manifestation of scintillations on the direction of daemon passage through the scintillator*

Now what could be the reason for the differences in the HPS width taking place as one crosses over from the high ($|\Delta t| \leq 20$ μs) to medium ($20 < |\Delta t| < 40$ μs) velocities, where they are insignificant, and to lower ones ($40 \leq |\Delta t| \leq 80$ μs), the situation where the upward moving daemon, on capturing a Zn or Sn nucleus, creates a longer scintillation observed by the upper PM tube than when it propagates downward?

The fact is that an HPS initiated by a daemon capturing a nucleus is, generally speaking, the result of two processes. The first of them is the fast emission of many energetic Auger electrons. They produce short NLSs. These NLSs are, however, immediately superimposed on by a series of true HPSs caused by evaporation from the nucleus of nucleons, likewise in large numbers.

When a slow enough ($40 \leq |\Delta t| \leq 75$ μs) upward-moving daemon (with $Z_{eff} = 1$) enters the bulk of the ZnS(Ag) layer and captures a nucleus, it emits tens of Auger electrons during $\boldsymbol{t}_{Aug}$. If $\boldsymbol{t}_{Aug} \ll t_l$, and the electron mean free path $\boldsymbol{l}_e \ll l$, the light produced in this NLS is also emitted from a layer of thickness $\approx \boldsymbol{l}_e \ll l$. Note that for a 25-keV electron, $\boldsymbol{l}_e \approx 1$ mg/cm$^2$, and only for a 50-keV one, $\boldsymbol{l}_e \approx 4$ mg/cm$^2$. Although some electrons are apparently emitted with energies of up to ~1 MeV, the maximum of their energy distribution may be expected to lie at low energies (possibly, at 1-10 keV). In order to reach the upper PM tube, the light from the lower part of the ZnS(Ag) layer has to cross practically the whole thickness ($l - \boldsymbol{l}_e$) of this layer. It is known that while the ZnS(Ag) phosphor is white, its transparency to its natural radiation is far from 100% because of scattering and absorption (Birks 1964). The subsequent HPS radiation generated by the daemon-containing evaporating nucleus propagating through the remaining scintillator layer (at $\boldsymbol{t}_{ev} \approx t_l$), which is closer to the PM tube, has more favorable conditions to reach the latter.

Now the daemons moving downward face the reverse situation. Here, the short NLS part of the scintillation is emitted in a thin ($\ll l$) layer facing the PM tube, while the longer HPS part, in the rest bulk of the light-scattering ZnS(Ag) layer. As a result, the total observed radiation will be, on the average, shorter even if it seems having the same amplitude.

For very slow daemons ($V \leq 5$ km/s), whose transit time $t_l$ through the ZnS(Ag) layer is substantially longer than $\boldsymbol{t}_{Aug} + \boldsymbol{t}_{ev}$, the situation of the scintillation shape changes again. In this case, the light from the scintillations excited successively by the Auger electrons and the evaporating nucleons should pass a practically the same path length $l$ in the ZnS(Ag) layer.

As a result, both scintillation components will suffer the same attenuation, and the difference in the paths passed by light in the bulk of the ZnS(Ag) layer can no longer be used to discriminate between the two, as was the case with the velocities ~10 ≤ V ≤ 35 km/s. Therefore, as both wings of the $N(\Delta t)$ distribution obtained for $\Delta t \approx 60\text{-}70$ μs become progressively more filled because of the slowing down of the GESCO daemons the differences between $N_w(\Delta t)$ and $N_n(\Delta t)$ first (for $V \to 5$ km/s) level off (see Fig. 4.2a,b), after which, for $|\Delta t| \geq 70\text{-}80$ μs ($V \leq 4$ km/s), $N_w(\Delta t)$ even surpasses $N_n(\Delta t)$ (Figs. 4.3a,b). This new reversal of the $N_w/N_n$ ratio occurring for $\Delta t \geq 70\text{-}80$ μs can be accounted for if both parts of a scintillation (NLS and HPL) are of about the same amplitude. In this case, the daemon scintillation will be slightly more filled than the standard background HPS, which is excited, for instance, by natural radioactivity. Obviously enough, the daemon-initiated scintillation will fall now into the right-hand part of the HPS Gaussian in Fig. 3. Fig. 5 displays the evolution of the average $N_w/N_n$ ratio for daemon scintillations as a function of $\Delta t$. Because of a slightly weaker attenuation of scintillations due to the downward-moving daemons, they will be detected in somewhat larger numbers than the scintillations caused by the daemons moving upward.

As seen from Table 1, in order for the scintillations to have the observed shape depending on the direction of the emitter motion in ZnS(Ag) (including the velocities of 35-50 km/s), the condition $t_{Aug} \leq 10^{-10}$ s $< t_l$ should be met, and the time of evaporative nucleus de-excitation should be $t_{ev} \approx 10^{-9}$ s (or slightly shorter). Because of the shape of the daemon-initiated scintillations depending on $\Delta t$, the available statistics are not yet sufficient to discriminate reliably between the HPSs caused by daemons and by the background.

### 4. On the Orbital Evolution of Daemons in Helio- and Geocentric Orbits

The fairly limited observational material presented above and its interpretation suggest already a certain pattern of the hierarchy of the near-solar daemon populations and of their possible evolution, which permits some predictions on the results which can be obtained in the forthcoming experiments.

(*i*) The population of the Galactic disk has a velocity dispersion less than or equal to the velocity of motion of the Solar system relative to the disk. The opinion (e.g. Freese et al 1988) that in the beginning of June the average ecliptical projection of the velocity vector of the Galactic halo population adds to the Earth's orbital velocity, and half a year later, in the beginning of December, is subtracted from it, is apparently applicable to a degree to the Galactic disk population too.

(*ii*) Because of the Galactic-disk daemons crossing the Sun being slowed down, the Sun captures some of them (with $V_\infty \leq 20$ km/s (Drobyshevski 1996)) into heliocentric orbits with perihelia lying inside it. Perturbations by the Earth transfer subsequently a small part of these daemons into the SEECHOs with perihelia outside the Sun (Drobyshevski 1997a). The SEECHO aphelia are oriented primarily in the direction of the "shadow" created by the Sun in the incoming flow of the Galactic disk population. The Earth crosses this shadow in March (and, possibly, in February as well). Possibly, a part of the already slowed down daemons, which, in the course of their capture by the Sun, falls on it back from the shadow region and, on traversing the Sun, exits it from the opposite side to enter the "antishadow", i.e., counter to the flow of the Galactic disk population, produces a weaker and diffuse "antishadow" SEECHO population.

(*iii*) Multiple interaction of the Earth with the SEECHO objects transfers a part of them gradually to NEACHOs (Drobyshevski 2000d). Because the SEECHOs concentrate primarily in the "shadow", the orbits of the NEACHO population should cross the Earth orbit (to be more precise, the torus swept by the Earth's sphere of influence) in the "shadow" region.

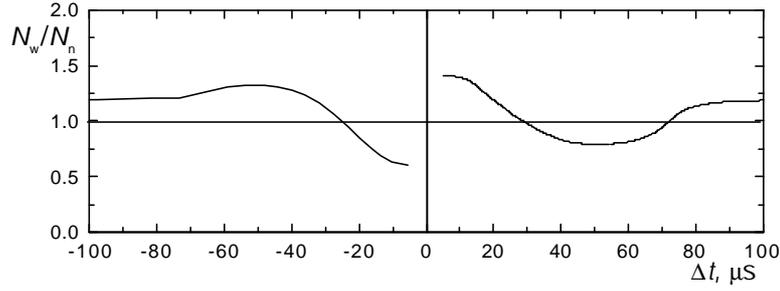

Fig.5. Schematic of dependence of the daemon-caused HPSs' asymmetry on the time-shift of the upper and lower scintillations $\Delta t$, i.e. of the daemon velocity (if taking the 29 cm distance between upper scintillator and the bottom case cover to be basic).

In this way a diverging beam of NEACHOs exiting the shadow region forms (a similar NEACHO focusing can occur in the region where the Earth's orbit crosses the antishadow). The daemons populating the NEACHOs fall on the Earth with a velocity ~11.2-15 km/s. It is this flux that we detected in March 2000.

(*iv*) As a daemon repeatedly meets the Earth and traverses it, its velocity relative to the Earth decreases to the extent where it can be captured into the GESCOs (Geocentric Earth-Surface-Crossing Orbits). It is this population that produced the maxima at ±50 μs observed in April 2001 (Fig. 4.1c,d). The fate of the daemons in GESCOs is clear; slowed down by the Earth, their velocity decreases gradually, and they drop below the Earth's surface. This scenario correlates fully with the spreadout and monotone displacement of the maximum in $N(\Delta t)$ from $|\Delta t| \approx 50$ μs in April to $|\Delta t| \approx 60$-$80$ μs thereafter, from the end of May to June (compare Figs. 4c). Thus, the average GESCO daemon velocity at the Earth's surface decreases by ~30-40% in a month. This figure offers a possibility to estimate the force with which the Earth's matter slows down the daemons. Bearing in mind that the capture of nuclei of the matter by the daemon may result in $Z_{eff} \nleftrightarrow 0$, an *ab initio* calculation of this force is in no way a simple problem. Now if we assume it to originate from momentum transfer to individual atoms (or their nuclei) in the daemon path, our experiments suggest that such interaction occurs, on the average, on a path $\boldsymbol{l}_{eff} \sim 10^{-6}$-$10^{-5}$ cm, which is much larger than the internuclear distances in matter (~$10^{-8}$ cm).

(*v*) Because the probability of nuclear capture and excitation of a scintillation in our ZnS(Ag) detector ~10 μm thick drops below unity for velocities $V > 35$ km/s, and, also, that $\boldsymbol{t}_{Aug}$ is comparable to $t_l$, it appears only natural to assume that we detect far from all events produced by daemon moving with $V > 35$ km/s. (Note that the system becomes transparent for particles with $V > 50$ km/s altogether.) Therefore to arrive at the lowest limit for the daemon flux through our detector, we may justifiably assume that the $-20 < \Delta t < 0$ μs bin in the inverted $[N_w(\Delta t) + N_n(-\Delta t)]$ distribution (Fig. 2d and Figs. 4d) may not practically contain daemon events, so that its level approaches that of the background. But then the background events ($\approx 20\pm5$ events per bin) constitute ~80-85% of the total number of all events. Whence the total downward flux in March 2000 is given by ~28 events, all of which belong to the $+20 < \Delta t < 40$ μs bin. Thus, the average downward flux is $f_\oplus \geq 10^{-5}$ m$^{-2}$s$^{-1}$. As already mentioned, the flux components vary in time.

The high-velocity component ($0 < \Delta t < 20$ μs), as evident from Fig. 2 and Fig. 4,

appears to present always, whence it follows that it is primarily due to the Galactic disk population. If it was the SEECHO population, we would also have always observed the NEACHO daemons. The fraction of the high-velocity component is small, apparently both because it does not build up in NEACHOs and GESCOs and, possibly, because of our system being transparent to daemons with high velocities ($\geq$40-50 km/s).

While the differences in the numbers of events for the inverted distributions $N_w(\Delta t) + N_n(-\Delta t)$ within the $-20 < \Delta t < 0$ µs and $20 > \Delta t > 0$ µs bins are seen by eye, they become statistically significant only when the events of Figs. 2d and 4d are summed over the whole four-month interval. One may assume the presence of seasonal variations; however, the data are still insufficient to reveal them.

The March distribution exhibits a flux with $V \approx$ 10-15 km/s ($+20 < \Delta t < 40$ µs). Tested by the $\chi^2$ criterion, the $N(\Delta t)$ distribution is not a constant level with a CL $\approx$ 99.99%. This flux is due to the NEACHO population. Some fraction of this population is captured into GESCOs, which becomes manifest in April in the appearance of two diffuse but statistically fairly significant maxima [see also the $N_w(\Delta t) + N_n(-\Delta t)$ sum in Fig. 4.1d] lying at $-80 < \Delta t < -40$ µs and $+40 < \Delta t < +80$ µs (the difference from $N$ = const has a CL = 95.8%).

Because of the gradual slowing down of the GESCO population, in the May $N(\Delta t)$ distribution these maxima spread out and shift toward larger $|\Delta t|$ to make the whole distribution within our $\Delta t = \pm 100$ µs limits filled well with a small dip at the center (the CL of its being not $N$ = const by the $\chi^2$ criterion is only 9%). Because the events belonging to the side maxima of N($\pm\Delta t$) displace gradually to the $\Delta t = \pm 100$ µs limits, the central minimum begins to spread out, with a noticeable maximum appearing at its center in the first half of June (Fig. 4.3c) (as a result, the CL of $N(\Delta t)$ being not $N$ = const starts again to increase and reaches 15% and even 85% in the second half of June, see Fig. 4.4c). The formation of this maximum may be associated with the daemon flux from the Galactic disk becoming maximum because of its velocity adding to the Earth's orbital velocity. However, this maximum could possibly be caused by a gradual decrease of the GESCO population velocity ($V \leq$ 4 km/s) (See Sec. 3.3.1). As a result, the 7 cm gap becomes basic because a time of its traversing increases up to value which is sufficent for the captured nucleus decaying down to $Z_n \leq 9$.

The net picture suggested by the totality of these events is that some of the daemons which crossed in March our 1 m$^2$ system with a time shift $+20 < \Delta t < 40$ µs are captured into GESCOs. In April through June, they only diffuse in the velocity space and decrease in the average velocity by ~30-40% per month.

## 5. Some Implications and Problems

Actually, all GESCO objects should eventually pass under the Earth's surface. The flux $f_\oplus \sim 10^{-5}$ m$^{-2}$s$^{-1}$ corresponds to $5 \cdot 10^9$ s$^{-1}$ daemons crossing the Earth's surface. Accepting $\approx$5000 s for the period of revolution along a GESCO, the total number of daemons trapped into the GESCOs in March and dropping gradually into the Earth will be ~$2.5 \cdot 10^{13}$. (If this process of trapping repeats in the solar antishadow half a year later, the latter figure doubles.) In 4.5 Byr, $2.5 \cdot 10^{13} \times 4.5 \cdot 10^9 \approx 10^{23}$ daemons have accumulated in the Earth, their total mass amounting to ~$2 \cdot 10^{18}$ g (Drobyshevski 2001). We are not going to discuss here their fate.

For the rate of the velocity decrease in the GESCOs found by us, most of them will escape under the Earth's surface half a year after the capture. Therefore at the end of summer, the GESCO aphelia are located near the Earth's surface, where the daemon velocity drops to very low levels. Considered in terms of the daemon hypothesis of the ball lightning (Drobyshevski 1997b), this would favor their formation, which may correlate with the highest frequency of their occurrence in this time of the year (~2/3 in July-August (Smirnov 1990)).

The daemon hypothesis has permitted us to construct an inherently uncontradictory and self-consistent scenario of the events occurring in our detector, which records, irrespective of details of the interpretation, the passage of some nuclear-active objects moving with low astronomic velocities.

We assumed them to be Planckian objects with a negative charge $Z \approx 10$. At the same time, we believed the daemon "poisoned" by the captured nucleus to start interacting with surrounding matter again as soon as its effective charge, due to the evaporation of the nucleus and the decay of its protons, becomes $Z_{eff} \geq 1$. However, because the daemon moves not in vacuum, and the probability for it to capture a nucleus from the air is fairly high, the time it spends with $Z_{eff} \geq 1$ is at best ~1 μs, which is much shorter than the time needed to evolve to $Z_{eff} \approx 1$ (~10 μs). The situation in polystyrene is still more severe; here the time in which the daemon captures a new nucleus at $Z_{eff} = 1$ is a few fractions of a ns. Therefore, first, we possibly detect not more than a tiny fraction of all objects, and, second, the detector response to the up/down direction should be extremely asymmetric (for daemons with $Z_{eff} \geq 1$ entering the ZnS(Ag) layer from air or polystyrene), whereas the actual asymmetry is quite moderate.

The solution to the problem lies possibly in the long time the captured light nuclei (C, N, O) having high excitation energies of the first level (~2-6 MeV) stay in the Rydberg levels. As a result, (i) the binding energy of a "poisoned" daemon to a nucleus is comparatively small, and (ii) the "poisoned" daemon does not touch the newly captured nucleus. If this is so, when the daemon enters a substance with heavy nuclei the light nucleus is shed off, the daemon captures the easily excitable heavy nucleus and enters it to initiate immediately its multi-stage disintegration, the process that accounts for the observed scintillations.

Another explanation is suggested by the above-mentioned daemon hypothesis of the ball lightning (Drobyshevski 1997b). By this hypothesis, to maintain the BL energetics at a level of 1-100 W, the fusion of two $^{12}$C nuclei should proceed at Rydberg levels with a rate of $10^{13}$-$10^{15}$ s$^{-1}$. This rate is determined primarily by those of the diffusion and capture by the daemon of $^{12}$C nuclei in carbon fibers. The reaction product $^{24}$Mg traverses in the solid a micron-scale path. The time during which the two nuclei stay in near-daemon Rydberg orbits until fusion can be only ~$10^{-18}$ s, so that the daemon resides in substances with a small $Z_n$ mostly in the "unpoisoned" (or slightly "poisoned") state. The only reason for our not detecting its passage from the products of the fusion catalysis of light nuclei is apparently their extremely small path length. This also excludes the possibility of producing a strong enough detectable scintillation which can be excited if many resultant nuclei enter the ZnS(Ag) layer during a short (~1 μs) time. The capture and behavior of light nuclei in the Rydberg levels, as well as the capture of heavier nuclei in these conditions, require a further study.

## 6. Conclusion

Attempts at interpretation of the processes occurring in our seemingly simple scintillation detector shows it to be in actual fact a fairly complex system. The detector was built specifically to reveal the passage through it of slow, electrically charged super-massive Planckian particles, the DM objects. In particular, the ideology underlying this system assumed from the outset the possibility of decay of a daemon-containing proton. Although the experiment has revealed a number of events that at first glance appeared strange, nevertheless most of them find a self-consistent qualitative interpretation within the framework of the daemon hypothesis and permit new conclusions on the properties and interaction of daemons with matter in general, and with the Sun and the planets, in particular.

It should be stressed that irrespective of our understanding of the finer details of the processes occurring on the atomic and subnuclear levels, the results of the experiment imply unambiguously the existence of a flux of fairly slow long-living microscopic nuclear-active objects, that possess a giant penetrating ability. It was assumed that it is these properties that would be characteristic of charged Planckian objects because of their huge mass.

The March 2000 observations revealed a daemon population in near-Earth, almost circular heliocentric orbits (NEACHOs). They showed the main information obtained by the detector to be contained in the extended scintillations, similar to those produced by $\alpha$-particles. The high-velocity population (35-50 km/s) assumed initially to exist in strongly elongated, Earth-crossing heliocentric orbits (SEECHOs) was not found at that time. After a one-year interruption spent to search for possible artifacts and errors in the measuring equipment, improvements of the detector, and the data processing techniques, and long control experiments, the observations were resumed at the very end of March 2001.

It was found that these new data no longer reproduce the results obtained in March 2000. However, a more comprehensive analysis taking into account the shape of the scintillations permitted us to reveal, in both previous and new data, clear indications of the existence of a high-velocity population, which may be related both to the SEECHOs and to a daemon flux from the Galactic disk (or their sum; detection in the future of seasonal variations in the intensity of these components will hopefully permit their discrimination). One detected also an extremely slow (3-5 km/s) population, which can exist only in geocentric, Earth-surface-crossing orbits (GESCOs). Further, the data accumulated in the April-June 2001 period show unambiguously a gradual evolution of the GESCO population toward a slowing down of their velocity by 30-40% per month, i.e., their gradually becoming confined within the Earth (which permits one, by the way, to estimate the force with which the material of the Earth decelerates the daemons). The fast evolution of the GESCO population with time offers an explanation for the absence in April 2001 of any indication of the NEACHO population, which was seen clearly in March 2000. The Earth passed apparently at that time through the "shadow" produced by the Sun in the incoming Galactic disk population. In this shadow should be located the aphelia of the SEECHO population, and here, near the Earth's orbit, cross the NEACHOs created by the repeated action of the Earth on the SEECHO objects. The passage of this zone, where the Earth transfers daemons from the NEACHO population to the GESCOs, can take little time, not over one to two months. In order to construct a final scenario of the orbital evolution of daemons and of their transfer from one population to another, round-the-year observations are naturally needed. It may, however, be conjectured even at this time that the passage in September through the weaker daemon "antishadow" may permit one to observe one more, antipodal, NEACHO population.

An analysis of the possible processes on the atomic and subnuclear levels initiated by daemons in the components of our detector permitted us to make numerical estimates of some parameters characteristic of these processes.

For instance, Auger electrons are ejected from a daemon-captured atom in $\sim 10^{-10}$ s, and their energies range from about 10 keV to $\sim 0.6$-1 MeV.

The de-excitation time of a nucleus captured by a daemon in ZnS(Ag) turns out to be somewhat shorter than $10^{-9}$ s. The characteristic excitation energies constitute several tens of MeV, and the excitation itself is removed by evaporation of nucleons and their clusters, which create extended scintillations.

Finally, we have refined the value of the decay time of a daemon-containing proton. It is $\Delta t_{ex} \sim 1$ µs. Because the daemon is a relativistic object of a Planckian scale, this quantity is significant for the construction and experimental testing of the future quantum gravity theory.